\documentclass[twocolumn,aps,showpacs,preprintnumbers]{revtex4}

\usepackage{graphicx}
\usepackage{dcolumn}

\begin{document}

\title{Software systems as complex networks: \\
structure, function, and evolvability of software collaboration graphs}
\author{Christopher~R.~Myers}
\affiliation{Cornell Theory Center, Rhodes Hall, Cornell University, 
Ithaca, NY 14853}


\begin{abstract}

Software systems emerge from mere keystrokes to form intricate
functional networks connecting many collaborating modules, objects, classes,
methods, and subroutines.  Building on recent advances in the study of
complex networks, I have examined software collaboration graphs
contained within several open-source software systems, and have found
them to reveal scale-free, small-world networks similar to those
identified in other technological, sociological, and biological
systems.  I present several measures of these network topologies, and
discuss their relationship to software engineering practices.  I also
present a simple model of software system evolution based on
refactoring processes which captures some of the salient features of
the observed systems.  Some implications of object-oriented design for
questions about network robustness, evolvability, degeneracy, and
organization are discussed in the wake of these findings.

\end{abstract}


\maketitle

\section{Introduction}
\label{Sec-Introduction}

In both the organic forms of nature and the engineered artifacts of
human society, complex systems grow and evolve to reveal intricately
networked organizations.  Surprisingly, the underlying structures of
these networks -- including the Internet~\cite{Faloutsos1999}, the
World Wide Web~\cite{Albert1999, Broder2000}, collaborations in
science~\cite{Newman2001a, Newman2001b} and cinema~\cite{Watts1998,
Barabasi1999}, interactions of proteins in yeast~\cite{Jeong2001}, and
metabolic pathways in a variety of organisms~\cite{Jeong2000} -- have
recently been found to share many ``scale-free'' and ``small-world''
qualities, which can be rather different from those found in simple
random networks.  These discoveries have served to draw together many
disparate fields into an emerging science of ``complex networks'',
which aims to unravel the principles by which networked systems form,
evolve, and remain robust and adaptable in the face of changing
environments~\cite{Strogatz2001, Barabasi2002}.

Software systems represent another important class of complex
networks, which to date have received relatively little attention in
this field.  Software is built up out of many interacting units and
subsystems at many levels of granularity (subroutines, classes, source
files, libraries, etc.), and the interactions and collaborations of
those pieces can be used to define networks or graphs that form a
skeletal description of a system.  Software systems are of course
important in their own right, as the centerpiece of the
information-based world in which we now find ourselves.  But they also
suggest some novel perspectives in the study of complex networks.
Especially important is the fact that software systems are organized
to be at once both highly functional and highly evolvable, with
evolvability often implemented through collective and collaborative
designs that target interfacial specificity as an important
controlling parameter.  This substantial emphasis on evolvability
makes software systems somewhat different from other engineered
systems, and closer in some ways to evolving biological systems.

Design is a central element of software construction, and many design
methodologies deal explicitly with the structure of software networks,
most often addressing the interactions of a small set of components at
a time. But the combined and persistent action of implementation,
refactoring, composition, extension, and adaptation leads to emergent
software organizations whose structures lie outside the realm of
explicit design.  Understanding the large-scale structural
organizations that form in software networks is increasingly important
not only for applications developed by distributed, loosely coupled
teams, but also for emergent computations that arise in adaptive,
self-organizing systems of autonomous computing agents.  More broadly,
understanding the functional organization of evolvable software
systems may provide models, metaphors, and tools to help us understand the
forces that serve to organize other classes of complex networks, whose
informational structures may not be as readily apparent.

The remainder of this section provides an overview of some relevant
aspects of software design, and describes in more detail the nature of
the collaboration networks examined here, specifically class
collaboration graphs and subroutine call graphs from several
existing open-source software systems.  Section~\ref{Sec-Results}
examines the structure of those collaboration networks, describes their
connection to other recently studied complex networks, and discusses
some of the software engineering implications of those observations.
Related work by others is discussed in section
\ref{Sec-Related}. In section~\ref{Sec-Refactoring}, I present a
simple model of evolving software systems based on refactoring
processes which captures some of the essential features of the
observed systems.  Section~\ref{Sec-complex-networks} highlights a few
issues suggested by the synergies among software systems,
object-oriented design, complex networks, and systems biology.
Section~\ref{Sec-Summary} provides a summary.

\subsection{Collaboration in software systems: function and evolvability}
\label{Subsec-collaboration}

Software engineering aims to decompose complex computational
functionality into many separate but interlocking pieces.  Rather than
reimplementing similar computations explicitly in every new context,
programmers develop abstractions of functions and data types that can
be used many times, in multiple contexts.  The ability to reuse
existing code can both speed the process of adding new functionality
(since new code need not be written) and facilitate the process of
modifying existing functionality (since changes can be made to a
single reused piece of code, rather than to multiple similar versions
of code strewn throughout a large application).  The process of
computation, therefore, involves collaboration: the distribution of
responsibility for computation among multiple software elements, such
as objects, classes, methods, subroutines, modules, and components.
These collaborations allow elaborate computational tasks to be built
up in a modular and hierarchical fashion, in loose analogy with the
way that sophisticated electronic circuits are built up from reusable,
low-level components.

Software collaborations imply dependency relationships, in that some
computational elements (e.g., classes and subroutines) need others in
order to carry out pieces of their appointed task.  One goal of
software design and development, therefore, is to construct an optimal
or near-optimal system of dependency relationships, whereby core
elements are reused in different contexts to perform recurring
fundamental tasks, with minimally constraining specializations added
in higher functional layers in order to build upon or combine those
fundamental tasks.  The utility of minimal specialization (typically
in the types of data being passed between subroutines and methods) is
tied to the goal of code reuse: software units that function only
under highly specialized conditions are generally less able to be
broadly reused than those units that require only as much
specialization as is required to implement a needed computation.  Many
factors can influence the decomposition that is actually chosen in a
particular software project, however.  For example, the process of
subdividing computations into minimal units and generating complexity
by combining those pieces does incur some overhead in computational
performance.  In scientific computing, the need for high performance
has historically outweighed other design concerns, leading to more
coarse-grained decompositions of data and functionality that enable
efficient numerical processing on large chunks of data unencumbered by
indirection and function calls.  Ongoing research in the field of
scientific computing aims to develop techniques for incorporating
advanced software design methodologies without incurring excessive
penalties in performance.

Building up software to carry out complex tasks is only one goal,
however.  The resulting system must also be \textit{evolvable}, that
is, transformable into a new system to accomplish new but related
tasks without excessive cost or disruption to the system as a whole.
The need to accommodate change is a major driving force in software
engineering; this might include changes in external user requirements,
underlying hardware platforms, forms of input data, or types of
algorithms used.  In poorly structured code, modifying or adding a
single feature may require updates to many files or subroutines, which
can themselves then cascade throughout the system.  To combat this,
many strategies have been developed to support the simultaneous
demands of function and evolvability, so that code modifications
remain localized.  Many of these strategies hinge on instituting
sufficient decoupling between subsystems, which enables developers to
avoid constraints and commitments that make it difficult to change one
part of a system without changing many others.

\textit{Design patterns}~\cite{Gamma1994}, for example, are
an important class of strategies to support software system
evolvability.  These patterns are motifs describing the relationships
among collaborating classes or objects in an object-oriented (OO)
software system which are effective at encapsulating variability and
change.  Different patterns identify different aspects of a system
which are likely to change; those aspects might include how objects are
created, which objects need to communicate with one another, or what
particular algorithms are used to solve a problem.  The highly variable
elements of a system are then encapsulated behind generic interfaces
or dedicated objects that act as brokers to mediate computational
activity, thereby decoupling objects so as to avoid excess
dependencies that can inhibit evolvability.  Design patterns and
related techniques are typically applied at a small scale, at the
microstructural level describing interactions among a few objects.  An
open question, therefore, is how microstructural design methodologies
conspire in the large to form macroscopic software structures.

Evolution and evolvability are of course central concepts in the
description of biological systems, different in important ways from
their meaning in software engineering.  While software engineering
involves intelligent action, biological evolution is blind, and does
not.  Nonetheless, from a systems perspective, there may be forms of
network organization which support adaptations that are applicable to
both blind biological evolution and intelligent software
design.  Furthermore, the software engineering community is
increasingly recognizing the value of prototyping and other forms of
interactive, trial-by-error design, in an effort to ``embrace change''
rather than struggle against the rapid pace of software
evolution~\cite{Beck1999}. Many of the software design patterns
mentioned above, which are now codified and part of a developer's
standard repertoire, were initially emergent and recurring solutions
that developers uncovered in their quest to build flexible, reusable
code that could operate in rapidly changing environments.

The distributed and collaborative nature of software design is
increasingly relevant at the social level of programmer interaction,
as well.  Many software projects begin as small efforts led by one or
a few people, only to grow into large activities involving many
developers scattered around the globe, a transition that has been
dubbed ``from the cathedral to the bazaar''~\cite{Raymond1999}.  While
organized design methodologies that support software growth and change
are useful even in the smallest projects, they become especially
important in distributed multi-developer efforts where many
individuals may work only on small pieces of the overall system.  The
work presented here does not explicitly examine the effects of
transformations in the social infrastructure of developers, but
examinations of that sort could prove quite interesting.

\subsection{Software collaboration graphs}
\label{Subsec-collaboration-graphs}

The interactions among software components are multidimensional and
multifaceted, and any representation of a software system typically
involves a slice or projection through that complex space of
interactions.  \textit{Call graphs} that describe the calling of
subroutines or methods by one another have long been used to
understand the structure and execution of procedural activity in
software systems, while \textit{class and/or object collaboration
diagrams} are used to glean insight into the relationships among
abstract data types in OO systems.  Static graphs typically describe
the set of interactions that are possible, while dynamic graphs
generated during program execution identify interactions that actually
take place under specific run-time conditions.  This work addresses
static class collaboration graphs arising in OO systems and static
call graphs arising in procedural systems, both of which can be parsed
from source code.


\begin{figure}
\includegraphics{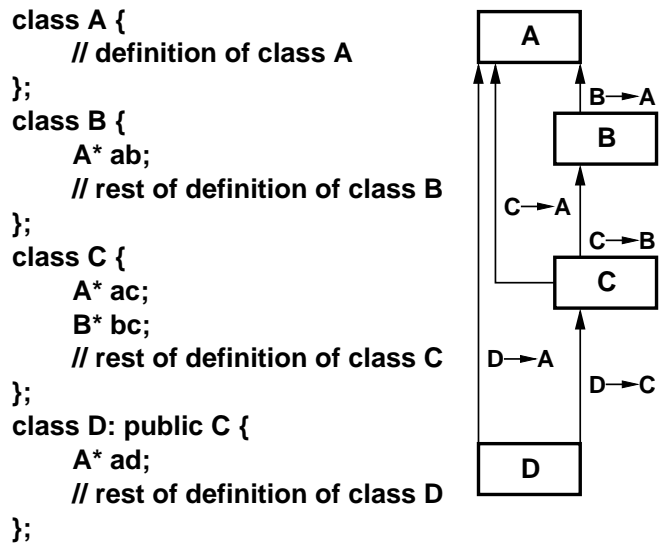}
\caption{\label{CollabExample}
Sample class collaboration graph (right) representing relationship
among classes \textbf{A}, \textbf{B}, \textbf{C} and
\textbf{D}, as specified by C++ class definitions (left).  Graph
nodes are classes, and graph edges are directed 
collaboration relationships between classes.  
This definition of class collaboration involves both 
aggregation relationships 
(\textbf{B}$\to$\textbf{A}, \textbf{C}$\to$\textbf{A}, 
\textbf{C}$\to$\textbf{B}, and \textbf{D}$\to$\textbf{A}) 
and inheritance/subclassing relationships
(\textbf{D}$\to$\textbf{C}).
}
\end{figure}


In OO systems, the definition of objects and their interactions plays
a central role.  Objects represent data types that are defined to
extend the basic, native data types provided by programming languages
(integers, floating point numbers, characters, etc.), in order to
develop more complex, application-specific, abstractions of data and
their associated behaviors (which are implemented via methods).
Typically, layers of objects are defined, building increasingly
complex representations by aggregating simpler ones.  An object
representing a vector field, for example, may combine simpler objects
representing vectors and spatial fields, which can themselves be
defined independently of each other.  (A vector object, for example,
might provide support for adding two vectors and computing their dot
product, whereas a spatial field object might support a
coordinate-based lookup to retrieve the value of the field at a
specified location.  Both vectors and spatial fields might themselves be
built up from even simpler objects, such as arrays.)  Classes describe
the form of objects in OO systems, and objects are instantiated at
run time from their class descriptions.

Class collaboration is the process by which more complex, multifunctional 
classes are built from simpler ones.  In this work, class
collaboration is defined to include the interaction of classes both
through \textit{inheritance} -- i.e., where one class is defined as a
subclass of another -- and through \textit{aggregation} -- i.e., where one
class is defined to hold an instance of another class.  A simple
illustrative example of such a graph is shown in
Figure~\ref{CollabExample}.  The direction of class collaboration and
subroutine calling follows standard software engineering convention
reflecting the flow of control in a system: an edge in a class
collaboration graph is directed from class \textbf{B} to class
\textbf{A} if \textbf{B} makes reference to \textbf{A} in its
definition (either through inheritance or aggregation), and an edge in
a call graph points from node \textbf{g} to \textbf{f} if subroutine
\textbf{g} calls subroutine \textbf{f} from within its scope.  The
definition of class collaboration used here allows one to decompose
the full class collaboration graph into two separate subgraphs, the
inheritance graph and the aggregation graph.


\begin{table}
\begin{tabular}{||l||r|r|r|r|r|r||}
\hline
           &   VTK &    DM &    AbiWord & Linux & MySQL &  XMMS \\
\hline
\hline
(a) & 788 & 187 & 1096 & 5420 & 1501 & 1097  \\ \hline
(b) & 1389 & 278 & 1857  & 11460 & 4245 & 1901  \\ \hline
\hline
(c) & 6 & 10 & 19  & 47 & 10 & 36 \\ \hline
(d) & 771 & 162 & 1035 & 5285 & 1480 & 971  \\ \hline
(e) & 1374  & 258 & 1798 & 11370 & 4231 & 1809 \\ \hline
\hline
(f)  & 4 & 2 & 46 & 10 & 12 & 0  \\ \hline
(g) & 5 & 6 & 25 & 6 & 7 & 0  \\ \hline
(h) & 8 & 10 & 72 & 9 & 10 & 0 \\ \hline
(i)  & 0.0165 & 0.0428 & 0.1332  & 0.0057 & 0.02 & 0.0 \\ \hline
\end{tabular}
\caption{\label{TableComponents}
Connected component analysis for the six software systems of interest.
Shown for each graph are: (a) number of nodes in each graph,
(b) number of edges in each graph, (c) number of WCCs, 
(d) number of nodes in the largest WCC, (e) number of edges 
in the largest WCC, (f) number of SCCs,
(g) number of nodes in the largest SCC, (h) number of edges in the largest
SCC, and (i) the fraction of nodes belonging to any SCC.
All six systems are characterized by a single 
dominant WCC, and little membership in SCCs.
}
\end{table}


I have examined collaboration networks associated with six different
open-source software systems.  These include class collaboration
graphs for three OO systems written in C++, and call graphs for three
procedural systems written in C.  The class collaboration graphs are
from: version 4.0 of the VTK visualization library~\cite{VTK}; the CVS
snapshot dated 4/3/2002 of Digital Material (DM), a library for
atomistic simulation of materials~\cite{DigitalMaterial, Myers1999};
and version 1.0.2 of the AbiWord word processing
program~\cite{AbiWord}.  The call graphs are from: version 2.4.19 of
the kernel of the Linux operating system, version 3.23.32 of the MySQL
relational database system, and version 1.2.7 of the XMMS multimedia
system.  Details on the construction and/or origin of these networks
is provided in the Appendix.

\section{Results}
\label{Sec-Results}

\subsection{Connected components}
\label{Subsec-Components}


\begin{figure}
\includegraphics{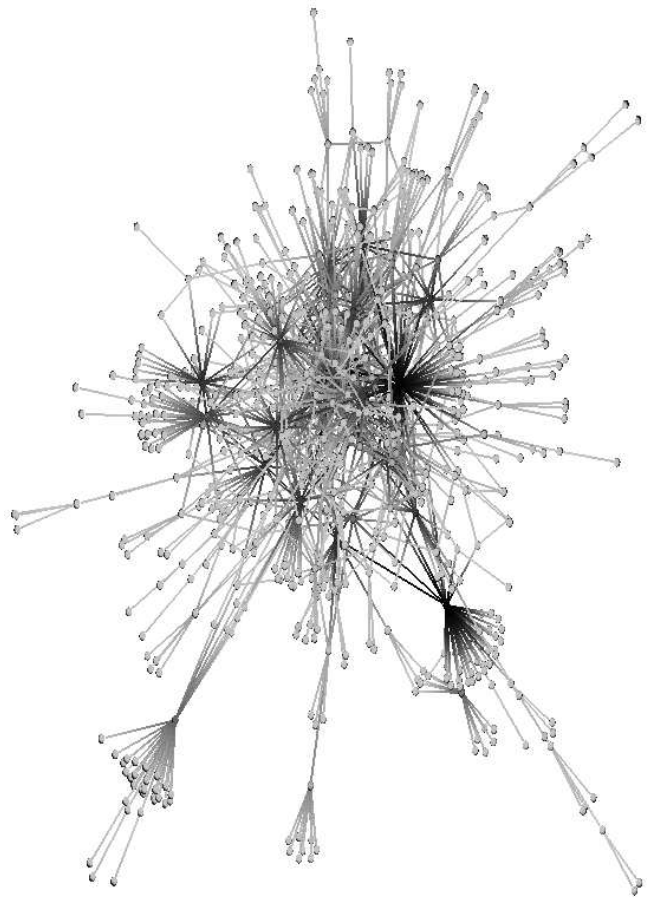}
\caption{\label{VTKgraph}
Largest weakly connected component of the class collaboration graph 
for the VTK system (layout courtesy of the Tulip graph visualization 
package).
}
\end{figure}


Connected components in a graph are those subgraphs for which all
nodes in the subgraph are mutually reachable by traversing edges in
the subgraph.  For a directed graph, one can define both weakly
connected components (WCC) and strongly connected components (SCC).
WCC are those connected components found in an undirected version of
the graph (i.e., by treating all edges as bi-directional), while SCC
are those connected components mutually reachable by traversing
directed edges.  By definition, every node in a directed graph will be
in some WCC; not all nodes, however, belong to an SCC.

Connected component analysis reveals trends across the six systems of
interest, which are summarized in Table~\ref{TableComponents}.  All
six systems consist of a single dominant WCC, comprising a large
fraction of the total nodes in the system (ranging from 86-99\%), and
a few (5-46) very small remaining WCCs.  A picture of the largest WCC
for the VTK system is shown in Figure~\ref{VTKgraph}.  Conversely, few
nodes belong to any SCC (less than about 4\% in five of the six
systems).  The lack of strong membership in SCCs is rather different
from that found in other directed complex networks, such as the
WWW~\cite{Broder2000} and various metabolic networks~\cite{Jeong2000}.
This difference is perhaps not surprising, given the nature of the
software graphs under investigation, which largely reflect aggregation
(of data in the case of the class collaboration graphs, or function in
the case of call graphs).  SCCs reflect subgraphs that are mutually
reachable, but the hierarchical directionality of use implicit in
software aggregation makes mutually reachable clusters of this sort
unlikely, and undesirable from a software development standpoint.
While generic base classes are reachable from their derived
subclasses, for example, the opposite is generally not true, since
that would undermine the genericity inherent in those base classes.
Similarly, one subroutine may call another as part of its execution,
but typically, the reverse is not true.

\subsection{Degree distributions}
\label{Subsec-Degree}

Degree distributions, summarizing the connectivity of each node in a
graph, are a feature that distinguish many complex networks from
simple random graphs.  For each node $i$ in a directed graph, there is
both an in-degree $k^{in}_i$, the number of incoming edges to node
$i$, and an out-degree $k^{out}_i$, the number of outgoing edges from
node $i$.  The in- and out-degree distributions $P^{in}(k)$ and
$P^{out}(k)$ indicate the probability of finding a node with a
specified in-degree or out-degree $k$, respectively, in a given graph.
Many complex networks have recently been found to possess a
``scale-free'' degree distribution~\cite{Barabasi2000}, indicating a
lack of characteristic scale (or degree) in the distribution $P(k)$.
This implies that $P(k)$ obeys a power law over an extended range of
degrees $k$: $P(k) \sim k^{-\gamma}$, or perhaps a power law truncated
by an exponential cutoff: $P(k) \sim k^{-\gamma} e^{-k/k_c}$.  In
contrast, a uniform random graph of $N$ nodes and ${\bar k}$ links on
average per node has a degree distribution with a characteristic scale
${\bar k}$, with $P(k)$ decaying exponentially away from ${\bar
k}$~\cite{Erdos1960}.


\begin{figure}
\includegraphics{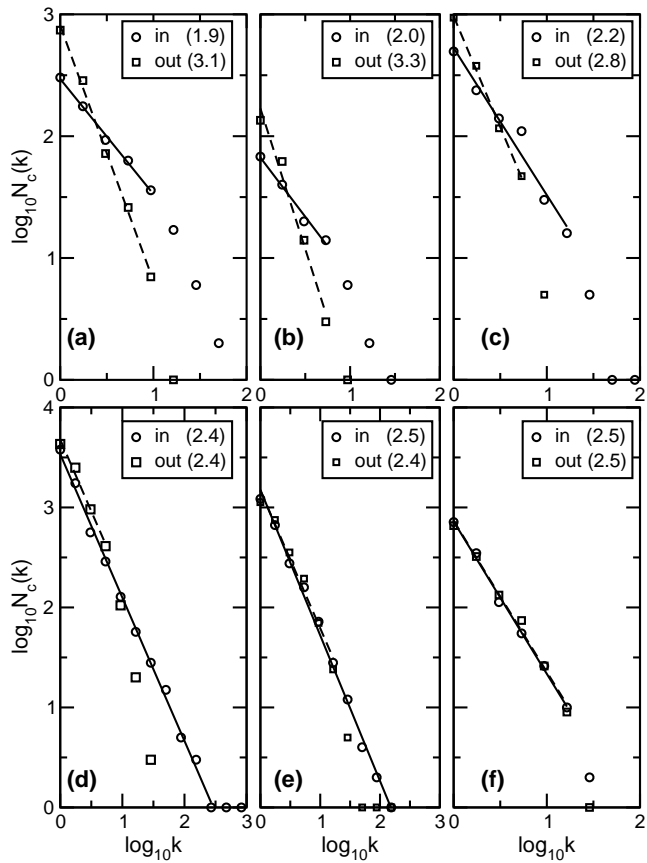}
\caption{\label{DegreeDistributions} 
Cumulative in-degree and out-degree distributions $N_c(k)$ for each of the 
six systems: (a) VTK, (b) Digital Material, (c) AbiWord,
(d) Linux, (e) MySQL, (f) XMMS.  $N_c(k)$ is the number of nodes in a 
graph with degree greater than or equal to $k$.
Histogram data for each distribution are
shown with symbols.  Lines indicate power-law fits (straight lines on
log-log scales) to histogram data in regions indicated; 
fits to in-degree distributions are represented by
solid lines, and fits to out-degree distributions by dashed lines.  The
legends show the values of the power-law exponents $\gamma^{in}$ 
and $\gamma^{out}$ for each fit (where $N_c(k) \sim k^{-\gamma + 1}$).
}
\end{figure}


I have examined the in- and out-degree distributions of the large
dominant WCC for each of the six software
systems of interest.  For each WCC, I have computed the unnormalized
cumulative frequency distributions $N_c^{\rm in}(k)$ and $N_c^{\rm
out}(k)$, where $N_c(k)$ indicates the number of nodes in a graph with
degree greater than or equal to $k$, and have plotted the logarithms
of these distributions in Figure~\ref{DegreeDistributions}.  ($N_c(k)$
is an unnormalized integral of the probability distribution $P(k)$;
for $P(k) \sim k^{-\gamma}$, $N_c(k) \sim k^{-\gamma + 1}$.)  These
distributions reveal a power-law scaling region (straight line on a
log-log plot), followed by a faster decay at large $k$.  
The extents of the power-law regions are admittedly small, 
particularly for the out-degree distributions, which one might
argue to be representative of exponential distributions.  Power-law
fits for all 12 distributions have been carried out over the regions
for which they exhibit scaling, and the values of the exponents
$\gamma^{in}$ and $\gamma^{out}$ are shown in the legends in
Fig.~\ref{DegreeDistributions}.

Interestingly, the class collaboration graphs shown in
Fig.~\ref{DegreeDistributions} reveal a marked asymmetry between the
in-degree and out-degree distributions, whereas the call graphs do
not.  In the class collaboration graphs, the out-degree exponent
appears to be significantly larger ($\gamma^{out}\approx 3$) than the
in-degree exponent ($\gamma^{in}\approx 2$).  For the procedural call
graphs, $\gamma^{in} \approx \gamma^{out} \approx 2.5$. For both sets
of graphs, the in-degree distributions tend to extend to higher $k$;
that is, it is more likely to find a node with many incoming links than
outgoing links.

As noted, the three class collaboration graphs also each contain an
embedded inheritance graph.  Since multiple inheritance is commonly
avoided (due to programming difficulties that it introduces), the
out-degree distributions of the inheritance graphs are strongly peaked
at $k=0$ and $k=1$.  The in-degree distributions for each of the three
inheritance graphs, on the other hand, also exhibit rough power-law
scaling, with exponents $\gamma \approx 2$ (not shown).  Therefore,
insofar as the in-degree distributions are concerned, the structural
forms of the overall collaboration graphs mirror those of the embedded
inheritance graphs.  The heavy tail in the out-degree distribution,
however, arises entirely from non-inheritance-based associations of
classes (such as the relationships 
\textbf{B}$\to$\textbf{A}, \textbf{C}$\to$\textbf{A}, 
\textbf{C}$\to$\textbf{B}, and \textbf{D}$\to$\textbf{A}
in Fig.~\ref{CollabExample}).

Classes and subroutines with small out-degree are generally simple,
since they do not aggregate other elements.  (If they became too
complex, there would be a pressure to break them up into simpler
pieces and introduce outgoing links.)  Conversely, elements
with large out-degree are generally more complex because they
aggregate behavior from many others.  Therefore, the existence of
heavy-tailed and/or scale-free out-degree distributions suggests a
broad spectrum of complexities.  On the other hand, classes and
subroutines with large in-degree are -- by definition -- reused in
many contexts, while those with small in-degree are not.  Thus, the
existence of heavy-tailed in-degree distributions implies a broad
spectrum of reuse.  Less interesting, perhaps, are those many
classes and subroutines that are neither heavily reused, nor heavily
constructed from other elements.

Software engineering practice encourages reuse, that is, large
in-degree; so it is not surprising that the largest degrees in these
systems are for incoming links.  It is not obvious, however, why the
class in- and out-degree distributions should be characterized by
quantitatively different scaling exponents.  The fact that the
procedural call graphs examined do not exhibit this in-out exponent
inequality also suggests further avenues of study: first and foremost
would be an examination of the call graphs associated with the three
OO systems studied, to ascertain whether in-out asymmetry is a
property of class collaboration or of object-oriented systems more
generally.

\subsection{Degree correlations}
\label{Subsec-degree-correlations}

\begin{figure}
\includegraphics{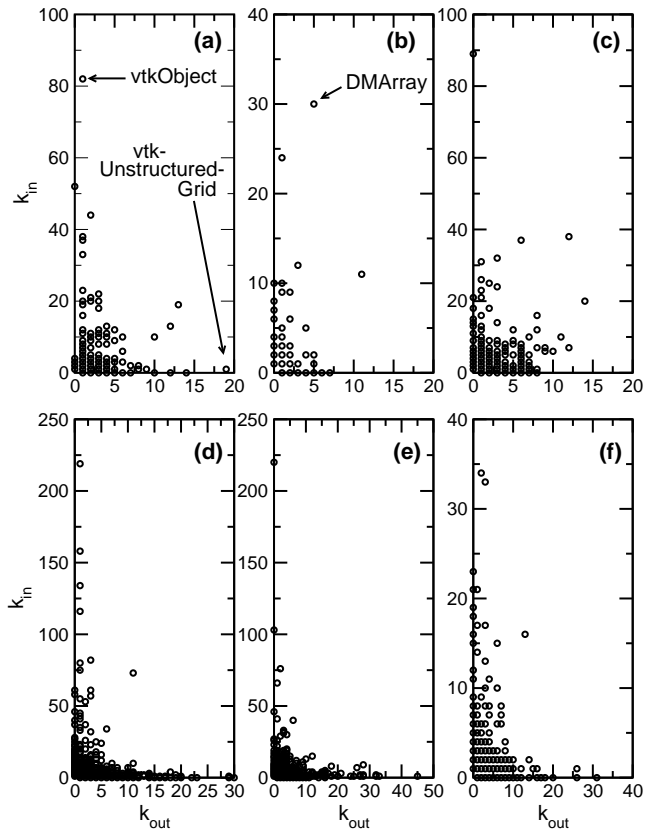}
\caption{\label{DegreeCorrelations} 
Scatter plot of the number of incoming links versus the number of outgoing
links, for every node in each of the six systems under consideration:
(a) VTK, (b) Digital Material, (c) AbiWord, (d) Linux, (e) MySQL, and (f)
XMMS.  
}
\end{figure}

Correlations among degrees can also provide insight into the structure
of complex networks.  The directedness of these graphs allows us to
examine the relationship between in-degrees and out-degrees.
Figure~\ref{DegreeCorrelations} shows this relationship for each of the
six systems of interest, where every node in each graph is represented
by its ($k^{out}, k^{in}$) pair.


It is visually apparent in Fig.~\ref{DegreeCorrelations} that nodes
with large out-degree generally have small in-degree, and those with
large in-degree have small out-degree.  This trend can be verified by
evaluating the linear (Pearson) correlation coefficient between the
sets $\{k^{in}\}$ and $\{k^{out}\}$ for those nodes with either a
large in-degree or large out-degree (or both); a threshold value of at
least ten edges has been chosen as a filter, somewhat arbitrarily as a
demarcation between the dense core of low-degree nodes and the sparser
set of high-degree nodes.  Table~\ref{TableDegreeCorrelations}(a)
shows the correlation coefficient for each of the six software systems
studied, demonstrating a negative correlation between in- and
out-degree for nodes with large degree in five of the six data sets.
Including the entire set of nodes for each data set (i.e., not
selecting only those nodes with $k \ge 10$) reveals, however, almost
no correlation in the call graph data (Linux, MySQL, XMMS), and weak
positive correlation in the class collaboration data (VTK, DM,
AbiWord).  This anticorrelation of large in- and out-degree implies
that, for the most part, there is a clear distinction between
large-scale producers of information (with high in-degree) and
large-scale consumers (with high out-degree).  Simple components tend
to be heavily reused, presumably because they are generic and
applicable to many different contexts, whereas complex components do
not, because they are highly specialized and only applicable in
limited contexts.

There are, however, outliers to that trend of separation; classes with
both large in-degree and large out-degree are evident in
Fig.~\ref{DegreeCorrelations}(a)-(c).  These classes have both
significant internal complexity (associated with aggregating the
behavior of several other classes), and significant external
responsibility.  There is reason to expect that such classes could be
problematic insofar as software development is concerned.  By way of
informal case study, my experience with development of the Digital
Material (DM) system (Fig.~\ref{DegreeCorrelations}(b)) confirms such
a suspicion.  The \texttt{DMArray} class identified implements both
array-like and tree-like functionalities, and is intended to be a
primitive data structure for much of the numerical computation in the
system.  Curiously, \texttt{DMArray} caused the most persistent
trouble within the DM development group.  The source of the difficulty
largely centered around the conflicting roles that \texttt{DMArray}
plays.  On the one hand, the class is intended to be a primitive black
box onto which more complicated functionality is to be layered, while
at the same time, it is itself a complicated datatype with substantial
internal structure and external behavior.  It would be interesting to
know whether other items with large in- and out-degrees apparent in
Fig.~\ref{DegreeCorrelations} have played similarly problematic roles
in the development of those various systems.  The complexity of
software components with large in- and out-degrees is highlighted in
the software metrics literature, for example, by the ``fan-in fan-out
complexity'' metric~\cite{Henry1981}, which states that complexity of
a code module is proportional to the square of the product of the
fan-in (in-degree) and the fan-out (out-degree) of the module.


\begin{table}
\begin{tabular}{||l||r|r|r|r|r|r||}
\hline
           &   VTK &    DM &    Abi & Linux & MySQL &  XMMS \\
\hline
\hline
(a) \textbf{In/out}  & & & & & & \\ \hline
\hline
$k \ge 10$ & -0.48 &  0.01 & -0.16  & -0.18 & -0.23 &  -0.75 \\ \hline
all $k$    & 0.09  &  0.10 &  0.18  & -0.01 & -0.03 &  -0.07 \\ \hline
\hline
(b) \textbf{Mixing} & & & & & & \\ \hline
\hline
in\ --\ in &  0.088 & -0.043 &  0.065 & -0.005 &  0.114 &  0.067 \\ \hline
in\ --\ out & -0.034 & -0.010 &  0.083 & -0.009 & -0.067 & -0.036 \\ \hline
out\ --\ in & -0.169 &  0.020 &  0.042 & -0.098 & -0.101 & -0.180 \\ \hline
out\ --\ out &  0.137 &  0.098 &  0.111 &  0.014 &  0.179 &  0.093 \\ \hline
\hline
undirected & -0.194 & -0.192 & -0.084 & -0.067 & -0.083 & -0.114 \\ \hline
\end{tabular}
\caption{\label{TableDegreeCorrelations}
Degree correlation coefficients.
(a) For each of the six software systems, 
correlation of $k^{in}$ and $k^{out}$ for each graph node,
for both the full data sets (``all $k$''), 
and for the reduced sets with 
$k^{in} \geq 10$ or $k^{out} \geq 10$.
(b) For each of the six software systems, 
degree mixing coefficients, relating in- and out-degrees for
each edge in the graphs.
The left column indicates the specific directed degree correlation function.
Also included is the correlation coefficient for an undirected 
version of each graph.
}
\end{table}


Another measure of degree correlation is the mixing by degree of a
graph~\cite{Newman2002}. This quantity measures the linear correlation
of degrees over all edges of a graph, i.e., the correlation of degrees
$k_i$ and $k_j$ for all sites $i$ and $j$ that define an edge in a
given graph, aggregated over all edges in that graph.  This reflects
the tendency of nodes of similar degree to be connected to one
another.  Most work on mixing by degree has focused on undirected
graphs or undirected versions of directed graphs, for which there is a
single correlation coefficient of interest: ${\rm corr}(k_i, k_j)$,
relating $k_i$ and $k_j$ for the node pair $(i,j)$ that are linked by
an edge in the graph. In a directed graph, there are four possible
correlation coefficients: ${\rm corr}(k^{out}_i, k^{out}_j)$, ${\rm
corr}(k^{out}_i, k^{in}_j)$, ${\rm corr}(k^{in}_i, k^{out}_j)$, and
${\rm corr}(k^{in}_i, k^{in}_j)$, where the index $i$ refers to the
source node of the directed edge, and $j$ refers to the destination
node.  Table~\ref{TableDegreeCorrelations}(b) shows the values of each
of these correlation coefficients, for each of the six graphs in
question.  Also computed is the degree mixing of the undirected
version of each software graph.  The undirected versions all show a
weak negative correlation (dissortativity) which suggests that nodes
with similar degrees tend not to be connected to each other.  The
directed graphs, however, tell a different story.  There, we find --
most significantly -- a weak positive correlation (assortativity)
among out-degrees, that is, a tendency for nodes with similar
out-degree to be connected.  There is a yet weaker positive
correlation among in-degrees.  While these correlations are rather
weak, their magnitudes are typical of those seen in a variety of other
complex networks~\cite{Newman2002}.

The weak positive assortativity among out-degrees seen in the software
networks is due in part to the hierarchical layering of functionality.
For example, in the VTK system, a complex aggregated class with large
out-degree such as the \texttt{vtkUnstructuredGrid} does not collaborate
directly with very low-level objects (such as
\texttt{vtkObject}); instead, it is built up out of collaborations
with ``mid-level'' objects (such as \texttt{vtkPolygon} and
\texttt{vtkHexahedron}), which themselves are aggregates built up from
lower-level classes.  Similarly, the weaker positive assortativity
among in-degrees for some of the graphs probably reflects the tendency
for simpler objects with large in-degree to collaborate with each
other at the base of a hierarchy of functional layers.  In the
undirected versions of these graphs, the negative correlation observed
reflects in part the fact that nodes with a large out-degree are
not linked to the nodes with a large in-degree, because of the
functional separation between producers and consumers described above.
These results further emphasize that the directedness of these graphs
is important.  Newman~\cite{Newman2002} has noted that sociological
networks (e.g., scientific collaborations) tend to be assortative
(positively correlated), but that technological and biological
networks are generally disassortative.  Clearly, for these software
graphs, teasing out the degree correlations requires examining
directed edges.

\subsection{Clustering and hierarchical organization}
\label{Subsec-clustering}

Clustering -- the tendency of a node's neighbors to be themselves
neighbors in the graph -- is a significant characteristic of
small-world networks.  While the results above indicated the
importance of graph directedness, clustering is typically measured on
undirected graphs.  For such a graph, the clustering coefficient $C_i$
of node $i$ is defined as $C_i = 2n/k_i(k_i-1)$, where $n$ is the
number of pairs of neighbors of node $i$ that are linked, and $k_i$ is
the degree of node $i$.  This quantity is simply the fraction of all
possible pairs of neighbors of node $i$ that are themselves linked in
the graph.


\begin{figure}
\includegraphics{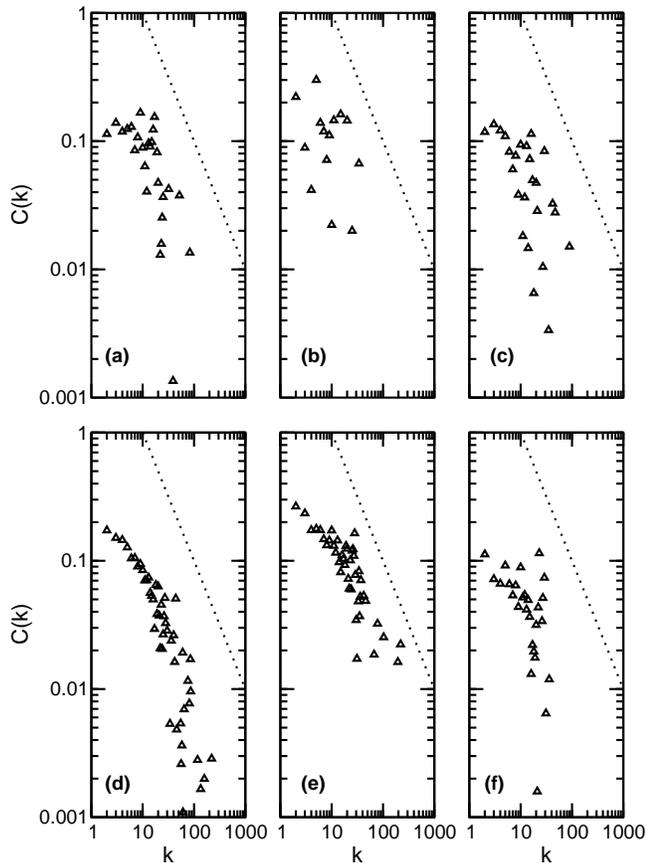}
\caption{\label{HierarchicalClustering}
Clustering coefficient $C(k)$ vs. degree $k$ for undirected versions 
of each of the six software systems.  $C(k)$ is the average clustering 
$C$ for all nodes in a graph with degree $k$.  Also shown for each graph is 
$k^{-1}$ scaling (dotted line), suggestive of hierarchical organization.  
Graphs are: (a) VTK, (b) Digital Material, (c) AbiWord, (d) Linux, 
(e) MySQL, (f) XMMS.
}
\end{figure}


Recent work by Ravasz \textit{et al.}~\cite{Ravasz2002a, Ravasz2003}
suggests that degree-dependent clustering of the form $C(k) \sim
k^{-1}$ is a signature of hierarchical organization in networks, and
can serve to distinguish hierarchical from non-hierarchical scale-free
networks.  They also suggest that hierarchical organization serves to
resolve the apparent dilemma between scale-free degree distributions
on the one hand (which imply no characteristic scale of connectivity)
and modular structure on the other (which suggest connectivities at
the scale of those consistent with modular units).
Figure~\ref{HierarchicalClustering} plots, for each of the six
software graphs of interest, the degree-dependent clustering $C(k)$,
defined as the average clustering $C$ for nodes in the undirected
graph with degree $k$.  These data are roughly consistent with those
presented in refs.~\cite{Ravasz2002a,Ravasz2003}, typically showing a
flat $C(k)$ for small $k$ which rolls over to a $k^{-1}$ tail at large
$k$ (more clearly defined for the larger graphs).  The $k^{-1}$ tail
is derived in refs.~\cite{Ravasz2002a,Ravasz2003} for a specifically
constructed hierarchical model; therefore, the existence of $k^{-1}$
scaling in real graphs would appear to be an indirect indicator of
hierarchical organization.  Nonetheless, given the hierarchical nature
of software design, further investigation of this sort of clustering
in software graphs is warranted.  Methods for extracting modules and
subsystems using the clustering data may also provide insight into the
organization of these systems~\cite{Ravasz2002a}.

\subsection{Topology, complexity, and evolution}
\label{Subsec-other-metrics}

Software systems can be characterized by a variety of metrics, which
can be compared to the underlying collaboration network
topology.  The VTK system, in particular, has been developed in a
manner that facilitates such measurements.  For every class in the VTK
system, I have calculated three quantities of interest: (a) total
source file size for each class; (b) total number of methods defined
for each class (including inherited methods), and (c) average revision
rate for each class over the lifetime of the VTK project (average
number of source file revisions per year, since initial commitment to
the VTK CVS source code repository).  Details on these calculations
are presented in the Appendix.


\begin{table}
\begin{tabular}{||l||r|r|r|r|r|r||}
\hline
VTK class complexity measures  &   in &    out \\
\hline
\hline
source file size & -0.28 &  0.58 \\ \hline
number of methods & -0.26 & 0.61 \\ \hline
average revision rate & -0.28 & 0.68 \\ \hline
\end{tabular}
\caption{\label{TableVTKCorrelations}
For VTK only, correlation of various class metrics (source file size,
number of methods, average revision rate) with in- and out-degrees.
}
\end{table}


For each class, these three metrics can be related to the in- and
out-degrees of that class; the linear correlation of these metrics
with degree is reported in Table~\ref{TableVTKCorrelations}.  We
see that all three metrics have a strong and positive correlation with
out-degree, and a weaker, negative correlation with in-degree.  Each
of these metrics reflects a different facet of class complexity
(implementation size, interface size, revision rate), and we see that
nodes with large out-degrees tend to be more complex than those with
large in-degrees, consistent with the scenario outlined previously.

\begin{figure}
\includegraphics{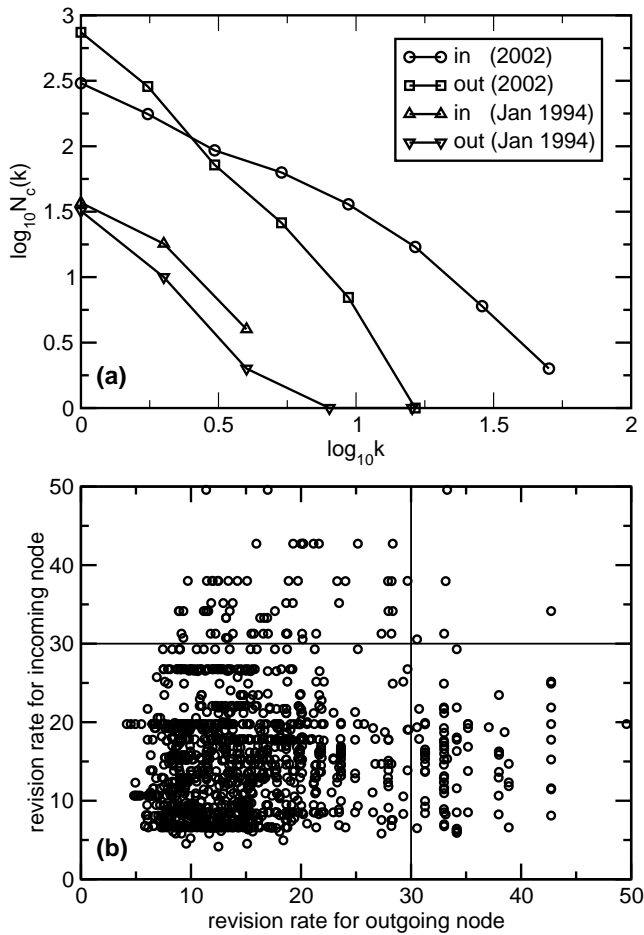}
\caption{\label{vtk_history} 
Facets of the evolutionary history of the VTK system.  (a) Comparison
of the degree distributions for the VTK system in its nascent state
(class collaboration graph as of Jan. 31, 1994), and for VTK version
4.0 (as presented in Fig.~\ref{DegreeDistributions}(a)).  (b) Scatter
plot of class revision rates for VTK collaboration graph.  Each point
represents an edge in the graph; the x- and y- coordinates of the
point are given by the average revision rates of the outgoing source and
incoming destination node, respectively, for that edge.  Note the
anticorrelation of revision rates for those classes with large rate
(e.g., greater than 30 per year).}
\end{figure}

The evolution of complex networks is a problem of great interest, and
the class revision history data for the VTK system provide some
insight into the evolutionary processes of software development, which
are summarized in Figure~\ref{vtk_history}.
Figure~\ref{vtk_history}(a) shows a comparison of the collaboration
graph degree distributions for the current VTK system (circa 2002)
with those for the system in its nascent state, at the end of January
1994 (the VTK ``reptile brain'', so to speak).  We see that the
heavy-tailed collaborative structure of the system was in place from
the outset, although the in-out asymmetry was less pronounced.
Figure~\ref{vtk_history}(b) examines the coevolution of classes that
collaborate in the VTK system.  For every edge in the VTK graph, the
revision rates of the outgoing source node and the incoming
destination node are plotted against each other.  From these data, we
find an interesting, and perhaps unexpected, trend.  For the entire
dataset, there is a weak positive correlation among revision rates for
connected classes (correlation = 0.10), indicating a weak tendency for
collaborating classes to evolve at the same rate.  But if we restrict
our focus to only those classes with large revision rate, e.g.,
greater than 30 revisions per year on average, we find a strong
negative correlation (correlation = -0.72).  This implies that
\textit{classes that evolve most quickly tend not to interact directly 
with each other}.

The data in Table~\ref{TableVTKCorrelations} reveal that, in the
VTK system, classes with large out-degree tend to evolve more rapidly
than do classes with large in-degree.  On the one hand, one might
imagine that the information consumers with large out-degree evolve
more rapidly simply because they have greater implementation sizes and
a larger number of methods subject to revision.  Alternatively, one
might imagine that evolution rate is primarily governed by the nature
of connections to other classes: information producers with large
in-degree are constrained to remain stationary (since other classes
depend on them), while information consumers are generally forced to
evolve to keep pace with changes in all the other classes that they
aggregate but are generally unconstrained by large numbers of users.
This latter scenario would suggest signatures of coevolution, that is,
correlation of evolution rates that are connected in the collaboration
graph.  

We saw in Fig.~\ref{vtk_history}(b) that there was a weak
coevolutionary tendency on average for the entire VTK system, but a
strong anticorrelation of coevolution among rapidly evolving classes.
The rapidly evolving classes are primarily (as seen in
Table~\ref{TableVTKCorrelations}) information consumers with
large out-degree, which are part of specialized functional subsystems
rather than the more generic functional substrate.  The strong
anticorrelation of interactions among highly evolving classes may thus
reflect a degree of modularity within the system, that is, the
functional separation of different specialized subsystems.  Whatever
weak coevolution there is appears to be confined to the more generic
substrate.  Further work is needed to explore the relationships among
evolution, connectivity, and constraint.  Similar explorations are
taking place within biology, to explore the relationship between
network connectivity (e.g., graphs of protein interactions) and
evolution rate~\cite{Fraser2002}.


\section{Related work}
\label{Sec-Related}

The fields of software metrics and reverse engineering examine
software systems in the aggregate, with significant emphasis on
measures of nodal degree in various software graphs.  By and large,
however, their scale-free nature appears to have escaped notice within
those communities.  Distributions of component connectivity are often
summarized in terms of means and variances, which are poor
characterizations of scale-free distributions.  The asymmetry between
large in-degree and large out-degree has long been identified by
software engineers as an important element of design, since large
out-degree indicates significant code reuse while large in-degree
indicates excessive object complexity~\cite{HendersonSellers1996}.

The existence of scale-free, small-world networks in software graphs
has been noted recently by a few groups, however.  Valverde 
and coworkers~\cite{Valverde2002, Sole2002} have examined class
collaboration graphs derived from the Java Development Framework 1.2
(JDK1.2) and the UbiSoft ProRally 2002 game system.  They have noted
scale-free degree distributions and larger-than-random clustering
characteristic of small worlds.  Their work does not distinguish,
however, between incoming and outgoing links in the software networks
they have studied; that is, they have formed undirected graphs by
ignoring edge directionality.  With undirected graphs, they could not
detect either the asymmetry between the in-degree and out-degree
exponents evident in Fig.~\ref{DegreeDistributions}, or the
anticorrelation of large in-degree and large out-degree seen in
Fig.~\ref{DegreeCorrelations}.  Perhaps more importantly, ignoring the
directionality of edges leads to different sorts of conclusions about
the implications of software engineering guidelines.  They claim that
software design is predicated in part on the ``rule of avoiding hubs
(classes with large number of dependencies, that is, large
degree)''~\cite{Valverde2002}, which does not recognize that large
out-degree and large in-degree have very different software
engineering implications.  Valverde \textit{et
al}.~\cite{Valverde2002}, however, do propose an interesting scenario
by which minimization of development costs might lead to an optimal
trade-off between developing a small number of large, expensive
components with few interconnections and a large number of small,
inexpensive components with many interconnections.  They suggest that
only sub-optimal solutions can be found in large, complex systems,
leading to scale-free/small-world behavior; this is an interesting
conjecture deserving further study, perhaps within the context of
synthetic models of software systems, such as the one that I will
introduce below.

Potanin \textit{et al.}~\cite{Potanin2003} have examined the structure
of object graphs, representing run-time snapshots of object
interactions in several OO programs.  Object graphs are the dynamic,
run-time analogs of the static class collaboration graphs studied
here.  Potanin \textit{et al.} observed power-law in-degree and
out-degree distributions, noting a tendency for in-degree exponents to
be clustered near 2.5, and out-degree exponents near 3, somewhat like
the trend that I have observed for class collaborations in this work.
They also note the strong separation of large in-degree and large
out-degree, similar to that presented here in
Fig.~\ref{DegreeCorrelations}.  Ultimately, developing a theory of the
relationships between static class graphs and dynamic object graphs
might prove useful to the software engineering community.

Wheeldon and Counsell~\cite{Wheeldon2003} have identified power-law
relationships in several OO measures, including inheritance and
aggregation graph degrees, and numbers of methods, fields, and
constructors defined by classes in OO systems.  They treat inheritance
and aggregation as two separate types of class collaboration (which
they are), while I have chosen to define collaboration more broadly to
include both.  They fit power laws to the full range of their
distribution data, even when there are apparent transitions between
scaling behaviors (e.g., from power-law to exponential), making
comparison with the present work difficult.  The numerical values of
the degree-distribution exponents they quote ($\gamma\approx 1$) are
rather different from those found here and in the work of Valverde
\textit{et al.}~\cite{Valverde2002} 
($\gamma\approx 2-3$), but it is not clear whether or not 
they have corrected
for the exponential bin sizes that they use to deal with sparse
statistics in the tails of the distributions.

Both Potanin \textit{et al.} and Wheeldon and Counsell make reference
to the process of preferable attachment as a mechanism for generating
scale-free networks, as outlined by Barab\'asi and Albert
(BA)~\cite{Barabasi1999}.  Preferential attachment was originally
proposed to describe the growth of the World Wide Web, but seems less
well-suited to describe the growth of software systems, although such
attachment is indirectly related in that low-specificity ``hubs'' will
generally attract more incoming links if they present broadly reusable
abstractions.  But the BA preferential attachment model is
acknowledged to be incapable of generating hierarchical structure,
which is clearly relevant for software design, and which is evident in
power-law, degree-dependent clustering such as seen here in
Fig.~\ref{HierarchicalClustering}.

Several features of the present research -- including examinations of
degree-dependent clustering, correlation of network topology with
various class complexity measures, and the evolutionary history of
class collaborations -- have not been explored in the other works
cited above.  Also, the present work introduces -- in the following
section -- a model of software evolution based on a set of standard
practices that captures some of the salient features of the observed
software networks.

\section{A refactoring-based model of software evolution}
\label{Sec-Refactoring}

As suggested earlier, software systems have a complex structure not
only to support the implementation of complicated functionality, but
also to allow for low-cost evolvability.  It is an interesting
question to ask, therefore, \textit{how software engineering practices
used to enhance system evolvability might alter the topological
structure of software collaboration graphs}.  An intriguing framework
for addressing such a question, and for generating models of evolving
software systems, is the set of processes collectively known as
\textit{refactoring}~\cite{Fowler1999}, which aim to remove
``bad smells'' from code that inhibit evolvability (e.g.,
extendability, modifiability, maintainability, and readability).
Refactoring tends to encourage the development of smaller, more
concise, single-purpose fragments of code (classes, methods, and
subroutines) that can be reused in a wider range of contexts, as
opposed to larger, multipurpose pieces of code that often contain
duplicated program logic.  Large methods and classes are often broken
up into collections of smaller ones, with appropriate indirection from
the former to the latter, leading to the creation of more nodes and
more edges in the resulting software graphs.  Duplicated pieces of
code are extracted from multiple locations in the source code, and
localized in a single place to which other pieces of code refer.  (In
some instances, however, a developer may deem such indirection
excessive and not worth the overhead; in those cases, refactoring
techniques would suggest the removal of nodes and/or the collapsing of
hierarchies.)  Many refactoring techniques can be cast in the language
of optimization, by minimizing (or altogether removing) bad smells
that pervade software.

Motivated by the basic observation that refactoring improves code
evolvability by reorganizing its internal structure, I have
implemented a simple model of an evolving software system, based on a
few refactoring techniques.  The model is overly simplified, insofar
as its treatment of software systems and practices is concerned, and
can never replicate the detailed structures of real software systems,
which are certainly history- and project-dependent.  Nevertheless,
some coarse features of the observed software networks can be
replicated with the simple refactoring model, which points the way
toward more sophisticated models of this sort, as well as further
empirical study of actual software systems undergoing refactoring.

In the model, binary strings of arbitrary length (i.e., strings
composed of $0$'s and $1$'s) serve as proxies for the subroutines in
call graphs and the classes in class collaboration graphs.  No attempt
is made here to distinguish between call graphs and class
collaboration graphs, so I will generically refer to these binary
strings as ``functions'' or ``strings'', and to the network of their
interactions as a ``call graph''.  Aggregation of functions is achieved
through concatenation of strings (represented here via the addition
operator).  Therefore, a larger, more complex, string can be built up
from smaller, simpler strings: e.g., $0110110010010 = 01101100 + 10010
= (01101 + 100) + (100 + 10)$.  Such a concatenation also has a
call-graph-based interpretation: the original parent node
($0110110010010$) has a link to each of the child nodes from which it
is composed ($01101100$ and $10010$), as do the second-generation
parent nodes to their children.  One can think of the original parent
node (before decomposition) as a single long function which calls no
other functions; after decomposition, the original function consists
only of calls to the new child nodes.  (Obviously, such a
decomposition is not unique, but that is true of software systems as
well.)


\begin{figure}
\includegraphics{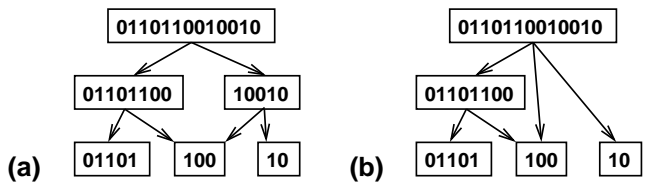}
\caption{\label{refac_model}
Processes implemented in the simple refactoring model.  (a) Long strings
are subdivided into two substrings, with links made to existing node
strings so as to avoid node duplication.   (b) One-off functions are 
removed to avoid excessive indirection; in this example, function $10010$ 
from part (a) is removed, 
and its parent is linked directly to its two children 
($100$ and $10$).
}
\end{figure}


In this model, three refactoring processes are implemented within
the framework of a Monte Carlo simulation, as schematically
represented in Figure~\ref{refac_model}: (1) functions that are
excessively long tend to be decomposed into a set of smaller
functions; (2) functions that already exist are used by others, rather
than having duplicated versions of the same function within the
system; and (3) some functions that are deemed to support ``excessive
indirection'' are removed, with appropriate rerouting of the call
graph.  More precisely, these processes are implemented as follows.
With probability $p$, a randomly selected function in the call graph
is decomposed into two smaller subfunctions with probability given by
$P(l) = {1 / (1 + {e^{-(l-l_0)/T}})},$ where $l$ is the length of the
string under consideration, and $l_0$ and $T$ are adjustable
parameters.  This probability is constructed in analogy with the
Fermi-Dirac distribution of statistical mechanics, and reduces in the
limit $T\to 0$ to the rule that any function with length greater than
$l_0$ will be decomposed (but which allows that threshold to be fuzzy
for non-zero temperature $T$).  A function that is selected for
decomposition is split into two subfunctions, with the breakpoint
selected at random (uniformly) anywhere in the string, as long as each
subfunction has at least unit length.  If either of the two
subfunctions already exists (i.e., is a defined node in the call
graph), a link from the parent function to that child subfunction is
created.  If a subfunction does not exist, a new node in the call
graph is created, and the parent is linked to it.  In this model,
therefore, no duplication of code is allowed (respecting the
proclamation of Beck and Fowler that duplicated code is ``Number one
in the stink parade''~\cite{Fowler1999}).  The final refactoring
process (removing ``excessive indirection'') is carried out with
probability $1-p$.  Specifically targeted are nodes in the call graph
that have only one parent (i.e., are called by only one other
function) and only two children; I will refer to such nodes as
``one-off functions''.  One such node from the set of eligible nodes
is chosen at random, and is removed from the system, such that its one
parent node is linked directly to its two child nodes.  This
specification is admittedly arbitrary, and could be further
parameterized; but the general purpose of such a process is to remove
functions that do not represent broadly useful abstractions (i.e., are
not used by many parent functions in different contexts) and that do
little more than to simply aggregate a small number of other functions
(i.e., two).

In the version of the model studied to date, the evolution process is
begun by constructing a call graph consisting of $N_0$ uniformly
random binary strings, each of length $L_0$, with no function calls
among them.  One could think of this initial set as a group of long
functions that are written explicitly in low-level code, with no
subroutines defined to abstract subunits of the computations.  As the
refactoring process unfolds, these overly long functions are
decomposed into sets of smaller functions, with links developing in
the call graph, and with smaller functions separated out for reuse by
others.  Initially, the only active refactoring processes are long
function decomposition and reuse of existing functions, since there
are no one-off functions to be removed at first.  Over time, however,
one-off functions become available for removal.  Such a system will
eventually reach an asymptotic steady state where all possible
refactorings have been carried out, although the decomposition of
strings with length $l < l_0$ can be very slow for small, nonzero $T$.
In the present work, I have stopped the refactoring process once the
size of the call graph ceases to change for at least 10000
consecutive refactoring steps.

\begin{figure}
\includegraphics{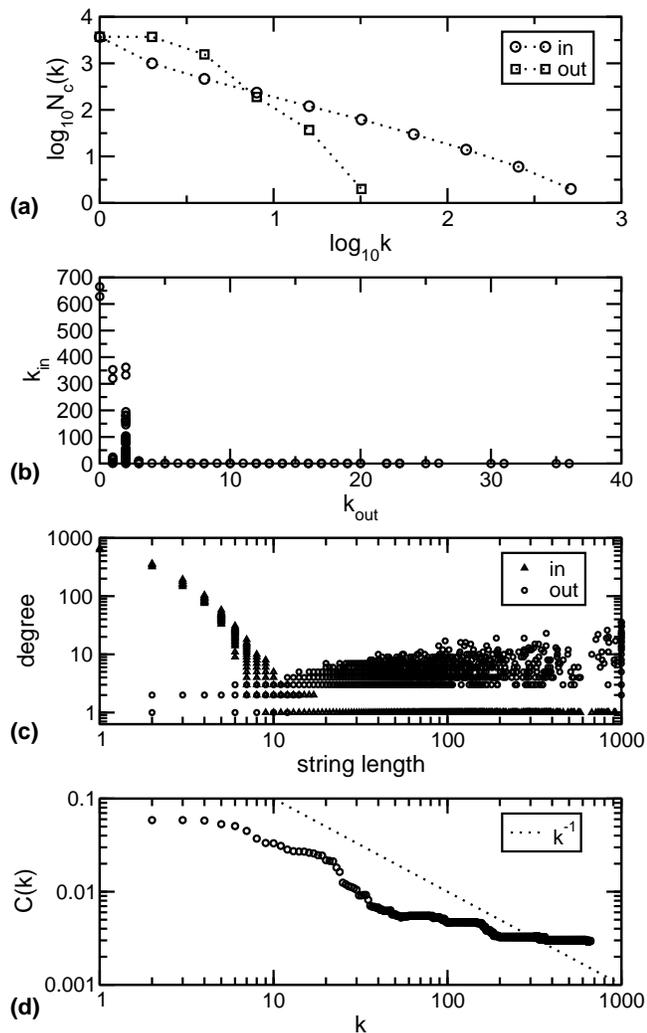}
\caption{\label{Refactory}
Simulation results for model of software evolution by refactoring: 
(a) in- and out-degree distributions, 
(b) scatter-plot of in- vs. out-degree,
(c) in- and out-degree versus string length, and 
(d) degree dependence of clustering, C(k) vs. k, also showing the 
$k^{-1}$ dependence suggestive of hierarchical 
organization.
Parameter values for the simulation were 
(see text for discussion): $N_0 = 50, L_0 = 1000,
p = 0.7, l_0 = 4, T = 1.0$.  
The initial graph with 50 nodes and 0 edges 
evolved into a graph with 3721 nodes and 14335 edges.
}
\end{figure}

Results from one such simulation are shown in Fig.~\ref{Refactory},
which plots the in- and out-degree distributions [part (a)], the in- vs
out-degree correlation [part (b)], the relationship between nodal degree
and string length [part (c)], and the degree-dependent clustering 
[part (d)] for the refactored software graph, in analogy with the results
presented in Figs.~\ref{DegreeDistributions},
\ref{DegreeCorrelations}, and
\ref{HierarchicalClustering}.


This simple refactoring model captures many of the salient features of
the observed systems.  Fig.~\ref{Refactory}(a) demonstrates
heavy-tailed degree distributions similar to those in the three class
collaboration graphs shown in Fig.~\ref{DegreeDistributions} (with
in-degree exponent $\gamma^{in} \approx 2$ and out-degree exponent
$\gamma^{out} \approx 3$), although -- as noted above -- there is
nothing specifically in the model that distinguishes class
collaboration graphs from call graphs.  A detailed examination of the
refactoring process reveals that the heavy tail with large in-degree
is generated by the process of large function decomposition and
associated reuse of existing functions (no code duplication).  The
tail with large out-degree arises from the process of one-off function
removal.  Fig.~\ref{Refactory}(b) reveals an in-out degree correlation
similar to that seen in Fig.~\ref{DegreeCorrelations}, with the large
in-degree and the large out-degree separated from one another.
Furthermore, the significance of these large degrees is similar to
what was identified previously, as is supported by
Figure~\ref{Refactory}(c), which plots nodal degree as a function of
string length for the resulting call graph.  Large in-degree/small
out-degrees represent small, simple functions (short strings) that can
be and, in fact, are reused in many contexts by many other larger
functions; whereas large out-degree/small in-degrees represent large,
complex functions that aggregate many smaller functions but which are
themselves used in few contexts.  Finally, Fig.~\ref{Refactory}(d)
reveals degree-dependent clustering similar to that suggested by the
hierarchical organization scenario and as seen in the data in
Fig.~\ref{HierarchicalClustering}, although the source of the
flattening of $C(k)$ at large $k$ is unknown.  As noted, the large
out-degree tail in Fig.~\ref{Refactory}(a) is generated by one-off
function removal.  Were this process not included, the out-degree for
every function would be either 0 (not decomposed) or 2 (decomposed
into two substrings).  Alternative rules for decomposing large strings
might produce heavy-tailed out-degree distributions even in the
absence of one-off function removal, but further work is needed to
explore in detail the behaviors of models inspired by refactoring
techniques.  Those interested in the tradeoffs between indirection and
computational performance might be motivated to construct a variant of
this model whereby long string decompositions incur a penalty cost
associated with performance degradation; other studies might try to
quantify the suboptimality suggested by Valverde \textit{et
al.}~\cite{Valverde2002} which could be associated with frustration of the
sort seen in other complex, disordered systems.

\section{Software systems and complex networks: the implications 
of object-oriented design}
\label{Sec-complex-networks}

\subsection{Robustness, fault tolerance, and evolvability}
\label{Subsec-robustness}

Researchers grappling to understand the structure, function and
evolution of complex networks highlight robustness as an important
theme in complex networks.  Studies have indicated~\cite{Jeong2001,
Carlson2002} that one by-product of scale-free networks in certain
systems is enhanced robustness in the face of random node failure
(although with increased fragility to failures of highly connected
nodes).  Analyses that highlight the similarities between biological
systems and engineering systems often emphasize redundancy in networks
as a mechanism for fault tolerance~\cite{Csete2002}.  Software
systems, on the other hand, are notoriously fragile, and at many
scales (e.g., single point mutations at the level of a typographical
error, inability to find a library needed for linking, cascading
modifications that follow from changes to a single class).  Of 
course, there are specialized applications which
absolutely require fault tolerance or are capable of exploiting
redundant resources in the case of network disruption (e.g., in
distributed processing applications).

By and large, however, the complexity that lurks within software
systems is not responsible for implementing fault tolerance and robust
control, as is the case for many complex engineering systems.  Rather,
much of the structural complexity of large software systems -- and in
object-oriented systems in particular -- is to support evolvability.
The need to continually accommodate and incorporate changes in the
external environment (user requirements, hardware platforms, etc.) 
lead to software designs that support modularity, decoupling, and
encapsulation~\cite{Gamma1994}. This connection between environmental
changes and adaptation toward modular network structure has been noted
in several other contexts, such as in models of biological
evolution~\cite{Lipson2002} and neural networks~\cite{Seth2003}.  In
contrast, very little of the complexity inherent in complex
engineering systems~\cite{Csete2002} is in place to support
evolvability and the construction of the next generation system
(despite the fact that design elements do get reused).

Much of the evolvability that is organized within software systems
arises from carefully planned genericity and associated decoupling,
using polymorphism and encapsulation to negotiate the inherent
trade-offs between specificity and evolvability of interactions.
While naive notions of object orientation suggest the proliferation of
increasingly deep class hierarchies that implement increasingly
specialized objects, the software engineering community has learned
that systems based on those sorts of objects and interactions are
often hard to modify~\cite{Gamma1994}. Design patterns aim to organize
the interactions of objects in such a way as to ensure sufficient
\textit{specificity for regulation and control} without unduly
freezing a system into \textit{commitments and constraints} that are
difficult to evolve.  Viewed in this manner, design patterns are
similar to biochemical processes such as regulated
recruitment~\cite{Ptashne2002} that serve to ensure specificity
through the cooperative action of several, more generic chemical
constituents, rather than the specific action of a single, complex
component.  It may be that scale-free network topologies help to
mediate the trade-offs between specificity and evolvability, and
present a mechanism for minimizing constraint while ensuring the
specificity required for regulation and control.

\subsection{Degeneracy and redundancy}
\label{Subsec-degeneracy}

The biological community has begun to make distinctions between
redundancy, involving the ability of identical elements to perform
identical functions, and degeneracy, involving the ability of
different elements to perform similar (or perhaps identical)
functions~\cite{Tononi1999, Edelman2001}.  This distinction emphasizes
the role that degeneracy can play in evolvability; whereas identical
and redundant elements are unable to provide any novel function in the
face of changing environments, similar and degenerate elements offer
avenues for adaptation because they offer the potential to provide
different behaviors in different contexts.  Degeneracy in biological
networks is in fact similar to polymorphism in object-oriented
systems, in that different objects can substitute for one another to
perform structurally similar functions which nonetheless differ in
detail.  This polymorphism imbues OO systems with evolvability by
enabling them to be more easily adapted to changing needs and
environments.  Tononi \textit{et al.}~\cite{Tononi1999} have developed
information-theoretic measures to quantify redundancy and degeneracy
in neuronal networks, and it is an interesting open question as to
whether connections between degeneracy and polymorphism might suggest
novel ways of analyzing and interpreting software systems based on
similar sorts of measures.

Curiously, Sol\'e \textit{et al.}~\cite{Sole2002} have commented that
software collaboration networks have ``a certain degree of
[redundancy] but no [degeneracy]''.  Their assertion of redundancy is
based on the existence of duplicated code, but this confuses
duplicated code with redundant code; duplicated code is not redundant
unless it is embedded in the same computational context, in the same
way that two identical resistors serving distinct roles in an
electrical circuit are not redundant to each other.  They claim that
there is no degeneracy in software because ``degeneracy is very common
[in] natural systems...but totally unknown with the context of
technological evolution''.  They further note that ``degeneracy is
intimately associated with tinkering in evolution: it reflects the
re-use that is made of different parts of a system in order to achieve
similar functions.''  I would argue that degeneracy is in fact quite
common in some software systems, largely in the form of polymorphism,
but perhaps through other mechanisms as well.  Furthermore,
generalizing from other technological systems to software systems is
problematic, in part because software is softer and more abstract than
other engineered systems.  And software design does itself involve a
significant amount of evolutionary ``tinkering'', which is becomingly
increasingly recognized and formalized through processes such as
refactoring~\cite{Fowler1999} and extreme programming~\cite{Beck1999}.

\subsection{Motifs, patterns, and emergent computational structures}
\label{Subsec-motifs}

There is growing interest in scanning large, emergent networks to
locate statistically significant, recurring motifs, and ultimately
identifying the functional significance of those
motifs~\cite{ShenOrr2002, Milo2002}.  Information processing systems
-- including gene transcription networks, neuronal systems, and
electronic circuits -- are seen to make use of recurring motifs such
as feed-forward loops and bi-fans (and in some cases, bi-parallel
subgraphs)~\cite{Milo2002}. A preliminary examination of recurring
motifs in the six software graphs studied here, using the motif
finding algorithm of Alon and co-workers~\cite{Milo2002}, identifies
these same motifs as being prevalent, but further study is needed to
examine their significance.  It remains to be seen, however, whether
such techniques will be able to identify meaningful software motifs
(e.g., design patterns) in all their glory, given the relatively crude
representations of software networks presented here.  The software
reverse engineering community is beginning to tackle the problem of
extracting complex design patterns from existing software systems, but
such work relies largely on detailed, \textit{a priori} specifications
of the structure of those patterns and more detailed class information
than is contained in the graphs studied here~\cite{Asencio2002}.  An
interesting challenge for the software engineering community would be
to develop systems and algorithms capable of extracting important
patterns and motifs from large software networks without such detailed
prior information.  Such an effort would not only be useful for
software design and analysis, but might also help to guide the field
of complex networks in identifying functionally important motifs.

One other interesting connection between software and complex networks
involves the very notion of ``software engineering''. As software
systems move increasingly into the realm of the emergent and the
unpredictable, a new notion of ``software science'' may unfold,
emphasizing fundamental phenomena to be explored, as opposed to cut
and dried systems to be built.  An interesting question concerns the
formation of structures akin to software design patterns.  In the same
way that recurring spatial patterns (vortices, dislocations, fronts,
and solitons) can arise in physical systems under stress, it may be
that recurring functional patterns (adapters, factories, mediators,
and proxies) can arise in appropriately defined computational systems
driven far from equilibrium.

\section{Summary and conclusions}
\label{Sec-Summary}

In this paper, I have examined several aspects of software
collaboration networks, inspired by questions in complex networks,
software engineering, and systems biology.  Not unlike findings by
others, the software collaboration networks studied all
exhibit scale-free and/or heavy-tailed degree distributions
qualitatively similar to those observed in recently studied biological
and technological networks.  An examination of these software systems
reveals that the hierarchical nature of software design has an impact
on the overall network topology.  Simpler, more generic classes and
subroutines form the heavy tail of the in-degree distribution, and
complex, more specialized aggregates populate the heavy tail of the
out-degree distribution, with the two generally well separated from one
another.  While the process of aggregation facilitates the
coregulation of many constituent elements, such control is also
constraining, and more difficult to evolve.  Design patterns,
polymorphism, refactoring, and related techniques aim to minimize
specificity of interactions while still enabling specific control, and
it may be that the scale-free nature of software collaboration
networks reflects these sorts of trade-offs in the large.

The work presented here highlights the need to preserve edge
directions in studies of directed software graphs, a fact that has
long been recognized within the software engineering community.  Edge
directionality is required to uncover several network features, such
as: differences between in- and out-degree distributions,
the anticorrelation between large in-degree and large out-degree, and the
positive assortative mixing among out-degrees.  Software collaboration
is inherently directed, and any attempt to explain network topologies
from software engineering principles or processes without recognizing
that asymmetry will most likely fall short.

More work is needed to better abstract and characterize the software
development process, and to uncover the implications of that process
for large-scale network topology.  Lehman's laws of software evolution
and their associated interpretation in the context of feedback and
self-regulation~\cite{Lehman1997} might form the basis of theories or
models aimed at uncovering large-scale structure from small-scale
process.  Further archaeology of existing software systems would also
help us to better understand relationships among network structure,
object complexities, object interactions, development processes, and
system evolution, and to unravel the differences between class
collaboration graphs and call graphs presented here.  Systems that
have undergone large-scale refactoring could be mined to ascertain
whether real-world refactoring processes change the nature of software
graphs (as they have in my simple model system), and large open-source
development projects such as Linux and AbiWord that have made
transitions from the ``cathedral'' to the
``bazaar''~\cite{Raymond1999} could be investigated for network-level
signatures of such transitions.

Combining insights from empirical studies of existing systems with
those gleaned from more abstracted models of software systems -- such
as the refactoring model presented here -- should be more fruitful
than either approach in isolation.  It would be interesting to learn
whether emergent, automatically generated computational systems,
such as those uncovered by genetic programming
techniques~\cite{Koza1992, Koza1994} or algorithmic
chemistries~\cite{Fontana1996}, give rise to the sorts of topologies
that are observed here.  It remains to be seen, however, whether
practical insights into the design and development of software
can arise from the consideration of software systems as complex
networks, more broadly construed.

Software systems present novel perspectives to the study of complex
networks.  Software is designed to be both functional and evolvable,
and those dual needs suggest particular forms of network organization.
Whereas other complex networks emphasize redundancy to support
fault tolerance, software networks highlight other degrees of freedom
that play a central role in supporting evolvability, such as
genericity, polymorphism, encapsulation, and collaboration.  If those
degrees of freedom are relevant to the organization and evolution of
biochemical networks, software systems may be useful in suggesting novel
insights into collective biological function.

\section*{Acknowledgments}

This work is supported by the USDA Agricultural Research Service under
project 1907-21000-009-04S (Specific Cooperative Agreement 58-1907-9-017).  
Resources of the Cornell
Theory Center (CTC) were used in this research; the CTC receives
funding from Cornell University, New York State, federal agencies,
foundations, and corporate partners.  I would like to thank David
Schneider, Sam Cartinhour, Jim Sethna, Josh Waterfall, Kevin Brown,
Jon Kleinberg, Shalev Itzkovitz, and Alex Iskold for useful
conversations, Maria Nemchuk for writing a web-crawler to extract VTK
CVS revision history data, Mel Gorman for making call graph data
available and clarifying its construction, and Uri Alon for making his
group's motif finding tool available.  I would also like to thank the
developers of the many software systems that contributed to this
effort: including: VTK, Digital Material, AbiWord, Linux, MySQL, XMMS,
Doxygen, Graphviz, CVS, Python, kjbuckets, NumPy, Scientific Python,
Grace, and Tulip.

\bibliography{softnet}

\appendix
\section*{Appendix: Materials and methods}
\label{Appendix-Materials}

The class collaboration graphs presented here (VTK, Digital Material,
AbiWord) are all based on subgraphs generated by
Doxygen~\cite{Doxygen}, an automatic document generation tool that
parses C++ header files to describe classes, their methods,
inheritance, and collaborations.  Doxygen generates a set of files
describing collaboration graphs in the ``dot'' format~\cite{Graphviz},
and Doxygen's definition of collaboration to include inheritance plus
association was used for this particular study.  Each class
collaboration graph is generated independently of all others, so all
the subgraphs must be assembled into a global collaboration
graph. Some minor typographic changes to class names were required to
enable programs in the Graphviz package to process the resulting graph
files.

The call graphs (Linux, MySQL, XMMS) were available for download on
the Web as demonstration data associated with the CodeViz package,
developed by Mel Gorman~\cite{CodeViz}.  CodeViz includes patches to
the gcc compiler that enable the extraction of static call graphs of
functions and macros.  As such, CodeViz does not include calls through
function pointers, nor does it capture inline functions or naming
collisions between functions with the same name in multiple files.

Connected component analysis was done using the \texttt{ccomps} and
\texttt{sccmap} tools in the Graphviz package~\cite{Graphviz}, for weak and 
strong connected component analysis, respectively.
As noted, subsequent analyses were 
carried out on the single, dominant weak connected component found
in each system.
Graph data and associated information are available online~\cite{Myers2003}.

The various class metrics for the VTK system presented in
section~\ref{Subsec-other-metrics} were extracted as follows.  Almost
every class in VTK is declared and defined in two separate source
files (header \texttt{.h} and implementation \texttt{.cpp}) whose
names correspond to the associated class.  Two exceptions are inner
classes (which are defined within their parent source files, but which
are identified by Doxygen as unique classes), and templated classes
(which in principle can produce multiple classes emanating from a
single pair of source files).  They are excluded from the analysis
relating class metrics to graph degrees, although they do contribute
to the source file sizes of their parent classes (introducing some
error to those sizes).  The computed source code file size for each
class is the sum of the total number of lines of code (including
comments and blank lines) in the two source files.  The number of
methods defined for each class are derived by combining Doxygen
information on class methods and the embedded inheritance graph
defined for each system.  Doxygen documents only those methods defined
within a class, which does not include methods inherited from base
classes.  The inheritance graph is thus traced to add to this list of
methods defined for each class those public methods defined by its
base classes.  Finally, because VTK has been developed within the
framework of the CVS source code revision system, information is
available describing the revision history of every source file in the
system (and hence, for each class in the system, because of the strict
mapping of classes to two source files).  This CVS revision
information is available on the Web~\cite{VTKCVSWeb}, which can be
crawled and parsed.  Any change to either the header or implementation
file of a class resulting in a new CVS version number for either of
those files was counted as a revision to the class; this, therefore,
counts twice any change to a method signature that would necessitate
an update to both the header and implementation files, leading to an
overestimation of rates, which may or may not be offset by the fact
that multiple source revisions could be swept under a single update to
the CVS system.  From these data, counts of the total number of
revisions of each class were generated, and divided by that class's
total lifetime in the CVS repository, to arrive at an average revision
rate since inception (expressed as average number of revisions per
year).  Classes that had been in the CVS repository for less than 3
million seconds (roughly 35 days) were excluded from the analysis,
since their short lifetime tended to introduce large errors in the
calculation of their revision rates.

\end{document}